\documentclass{aa}  

\usepackage{hyperref}
\hypersetup{
    colorlinks=true,
    linkcolor=blue,
    filecolor=blue,      
    urlcolor=blue,
    citecolor=blue
}
\bibpunct{(}{)}{;}{a}{}{,}
\usepackage{graphicx}
\usepackage{txfonts}
\defcitealias{dorman15}{D15}
\defcitealias{san12}{San+12}
\defcitealias{Kw12}{Kw+12}
\begin{document}

   \title{The survey of Planetary Nebulae in Andromeda (M31)} \subtitle{II. Age-velocity dispersion relation in the disc from Planetary Nebulae} 
    \titlerunning{The survey of PNe in M31 II - Imaging}
    \author{Souradeep Bhattacharya\inst{1}\and
          Magda Arnaboldi\inst{1} \and 
          Nelson Caldwell\inst{2} \and
          Ortwin Gerhard\inst{3} \and
          Mat\'ias Bla{\~n}a\inst{3} \and
          Alan McConnachie\inst{4} \and
          Johanna Hartke\inst{5} \and
          Puragra Guhathakurta\inst{6} \and
          Claudia Pulsoni\inst{3} \and
          Kenneth C. Freeman\inst{7}
          }
   \institute{European Southern Observatory, Karl-Schwarzschild-Str. 2, 85748 Garching, Germany \\ 
   \email{sbhattac@eso.org} \and
   Harvard-Smithsonian Center for Astrophysics, 60 Garden Street, Cambridge, MA 02138, USA \and
   Max-Planck-Institut für extraterrestrische Physik, Giessenbachstraße, 85748 Garching, Germany \and
   NRC Herzberg Institute of Astrophysics, 5071 West Saanich Road, Victoria, BC V9E 2E7, Canada \and
   European Southern Observatory, Alonso de C\'ordova 3107, Santiago de Chile, Chile \and
   UCO/Lick Observatory, Department of Astronomy \& Astrophysics, University of California Santa Cruz, 1156 High Street, Santa Cruz, California 95064, USA \and
   Research School of Astronomy and Astrophysics, Mount Stromlo Observatory, Cotter Road, ACT 2611 Weston Creek, Australia 
             }

   \date{Received: May 16, 2019/ Accepted: September 20, 2019}

 
  \abstract
   {The age-velocity dispersion relation is an important tool to understand the evolution of the disc of the Andromeda galaxy (M31) in comparison with the Milky Way.}
   {We use Planetary Nebulae (PNe) to obtain the age-velocity dispersion relation in different radial bins of the M31 disc.}
   {We separate the observed PNe sample based on their extinction values into two distinct age populations in the M31 disc. The observed velocities of our high- and low-extinction PNe, which correspond to higher and lower mass progenitors respectively, are fitted in de-projected elliptical bins to obtain their rotational velocities, $V_{\phi}$, and corresponding dispersions, $\rm\sigma_{\phi}$. We assign ages to the two PNe populations by comparing central-star properties of an archival sub-sample of PNe, having models fitted to their observed spectral features, to stellar evolution tracks. }
   {For the high- and low-extinction PNe, we find ages of $\sim2.5$ Gyr and $\sim4.5$ Gyr respectively, with distinct kinematics beyond a deprojected radius R$\rm_{GC}= 14$ kpc. At R$\rm_{GC}$=17--20 kpc, which is the equivalent distance in disc scale lengths of the Sun in the Milky Way disc, we obtain $\rm\sigma_{\phi,~2.5~Gyr}= 61\pm 14$ km s$^{-1}$ and $\rm\sigma_{\phi,~4.5~Gyr}= 101\pm 13$ km s$^{-1}$. The age-velocity dispersion relation for the M31 disc is obtained in two radial bins, R$\rm_{GC}$=14--17 and 17--20 kpc.}
   {The high- and low-extinction PNe are associated with the young thin and old thicker disc of M31 respectively, whose velocity dispersion values increase with age. These values are almost twice and thrice that of the Milky Way disc stellar population of corresponding ages. From comparison with simulations of merging galaxies, we find that the age-velocity dispersion relation in the M31 disc measured using PNe is indicative of a single major merger that occurred 2.5 -- 4.5 Gyr ago with an estimated merger mass ratio $\approx$ 1:5. }


   \maketitle
%

\section{Introduction}
Discs in late-type galaxies contain two distinct dynamical populations, the ``cold'' thin disc and the ``hot'' thick disc, as found in the Milky Way \citep[MW; e.g.][]{Gilmore83} and nearby edge-on galaxies \citep{Yoachim06, Comeron19}. The thin disc stars are younger with rotational velocities close to that of the collisional gas \citep{Roberts66}. The thick disc stars are older and through dynamical heating, via secular evolution of the disc \citep{Sellwood14} or mergers with satellite galaxies \citep{Quinn86}, their rotational velocity decreases and velocity dispersion increases. In the solar neighborhood, velocity dispersion appears to increase monotonically with age \citep{Delhaye65, Casagrande11}, but it is not known if that is representative of the entire MW disc. A major part of the MW thick disc is also found to be chemically distinct from its dominant thin disc \citep[see review by][]{bhg16}. While mergers of satellite galaxies can dynamically heat galactic discs, the cold MW disc does not seem to have undergone any major merger event in the last 10 Gyr.

The Andromeda (M31) galaxy is the closest giant spiral galaxy to the MW. A number of substructures have been observed in the inner halo of M31 \citep{mcc18}, which may have resulted from a recent merger \citep{Fardal13,ham18}. This may also be linked to an observed burst of star formation $\sim2$ Gyr ago \citep{bernard15,wil17}. It is well known that the velocity dispersion of a disc stellar population increases with age \citep{Stromberg25,Wielen77}.
\citet[][hereafter D15]{dorman15} estimated the age-velocity dispersion relation (AVR) for the M31 disc with kinematics of stars from the SPLASH spectroscopic survey \citep{Guha05, Guha06}. They assigned ages to stars based on their position on the colour-magnitude diagram (CMD, see their Figure 6) from the PHAT survey \citep{dal12}. While their main-sequence stars (MS) are well separated in this CMD, the red giant branch (RGB) and asymptotic giant branch (AGB) loci overlap, resulting in a more ambiguous age separation. They construct the line-of-sight (LOS) velocity dispersion ($\rm\sigma_{LOS}$) vs radius profiles for the different populations. Their Figure 16 shows a general trend of increasing $\rm\sigma_{LOS}$ with age, although with significant overlap of these profiles among populations and radii. On the basis of this trend, they state that the RGB population in M31 has a velocity dispersion that is nearly three times that of the MW. 


Planetary Nebulae (PNe) are discrete tracers of stellar populations and their kinematics have been measured in galaxies of different morphological types \citep[e.g.][]{coc09, cor13, pul18, Aniyan18}. PNe in M31 have negligible contamination from MW PNe or background galaxies \citep{Bh+19}. A number of PN properties are related to the mass, luminosity and ages of their progenitor stars. For example, from the central star properties derived from modelling nebular emission lines of PNe in the Magellanic clouds and M31, \citet{Ciardullo99} find a correlation between PNe circumstellar extinction and their central-star masses. Dust production of stars in the AGB phase scales exponentially with their initial progenitor masses for the $1\sim2.5 M_{\odot}$ range after which it remains roughly constant \citep{Ventura14}. Additionally, PNe with dusty high-mass progenitors evolve faster \citep{MillerB16} and so their circumstellar matter has little time to disperse, while PNe with lower central star masses evolve sufficiently slowly that a larger fraction of dust is dissipated from their envelopes \citep{Ciardullo99}. Kinematics of young and old stellar populations are well-traced by high and low mass giant stars respectively in the MW through their rotational velocity and velocity dispersion \citep[e.g.][]{Aniyan16}. In the M31 disc, different kinematics of younger and older stellar populations are expected to correlate with high- and low-extinction PNe respectively. 

\begin{figure}[t]
	\centering
	\includegraphics[width=\columnwidth,angle=0]{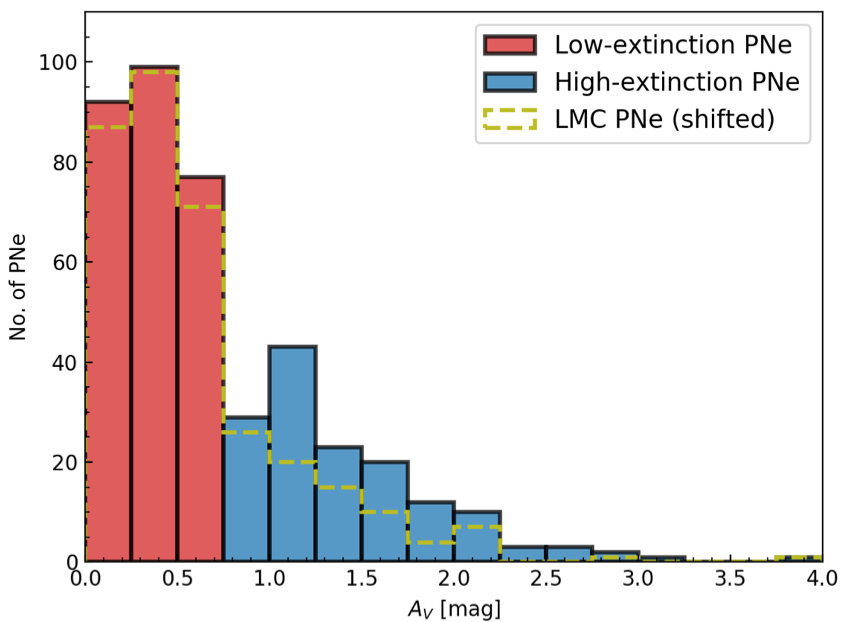}
	\caption{The histogram showing the distribution of extinction values for the \citetalias{san12} and Bh+19b PNe. The high- and low-extinction PNe lie in the blue and red shaded regions respectively. The distribution of extinction values of the LMC PNe observed by \citet{rp10}, shifted such that its peak corresponds to that of the M31, are shown in yellow.}
	\label{fig:av_hist}
\end{figure}

\begin{figure}[t]
	\centering
	\includegraphics[width=0.9\columnwidth,angle=0]{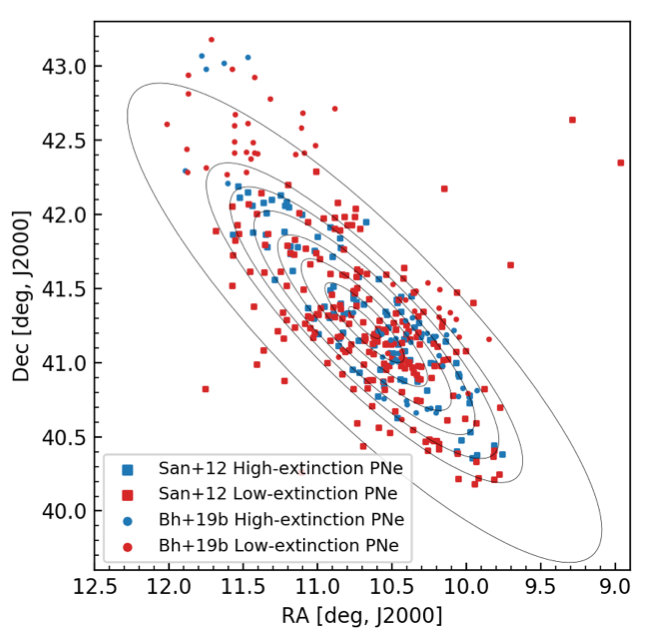}
	\caption{Position on sky of \citetalias{san12} PNe (squares) and Bh+19b PNe (circles). The high-extinction PNe are shown in blue while the low-extinction PNe are shown in red. The PNe are divided into the elliptical bins to obtain rotation curves.}
	\label{fig:spat_pos}
\end{figure}

In this letter, we identify PNe populations based on their extinction for the first time, allowing us to get two distinct age populations for the M31 disc. The data used in this work are described in Section~\ref{sect:data}. In Section~\ref{sect:ana}, we first discuss classification of PNe based on their extinction values. We then obtain the rotational velocity curve and rotational velocity dispersion for the M31 disc high- and low-extinction PNe. We assign ages to the two PNe populations by comparing modelled central star properties in \citet[][hereafter Kw+12]{Kw12} to the \citet{MillerB16} stellar evolution tracks. We then obtain the AVR for the M31 disc in Section~\ref{sect:avr} and compare it with previous determinations in M31 and the MW. From the comparison with simulated galaxies, we estimate the mass ratio of a possible merger event in the M31 disc. We summarise our results and conclude in Section~\ref{sect:sum}.


\section{Data description}
\label{sect:data}
\citet{Bh+19} identified PNe candidates in a 16 sq. deg. imaging survey of M31 with MegaCam at the CFHT, covering the disc and inner halo. Spectroscopic observations of a complete subsample of these PNe candidates were carried out with the Hectospec multifiber positioner and spectrograph on the Multiple Mirror Telescope \citep[MMT;][]{fab05}. The Hectospec 270 gpm grating was used and provided spectral coverage from 3650 to 9200 $\AA$ at a resolution of $\sim5~\AA$. Each Hectospec fiber subtends 1.5$\arcsec$ on the sky and were positioned on the PNe candidates in each field. On September 15, 2018, and October 10, 2018, with an exposure time of 9000 seconds each, two fields in the south-west region of the M31 disc were observed, while on December 4, 2018, with an exposure time of 3600 seconds, one field covering the northern part of the M31 disc and the Northern Spur substructure was observed. Of the 343 observed PNe candidates, 129 had confirmed detection of the [\ion{O}{iii}] 4959/5007 $\AA$ emission-lines. Of these observed PNe, 92 showed the H$\beta$ line and their extinctions (A$_V$) could be determined from the Balmer decrement. Details of the spectroscopic observations of the PNe, along with the extinction determination and chemical abundances will be presented in a forthcoming paper (Bhattacharya et al. 2019b in preparation, hereafter Bh+19b). 

\citet[][hereafter San+12]{san12} also studied PNe and \ion{H}{ii} regions in the M31 disc and outer bulge using Hectospec on the MMT. They observed 407 PNe, 321 with the H$\beta$ line detected and subsequent reliable extinction measurements. The combined sample of PNe with extinction measurements in M31 from \citetalias{san12} and Bh+19b thus consists of 413 PNe. 


\begin{figure}[t]
	\centering
	\includegraphics[width=\columnwidth,angle=0]{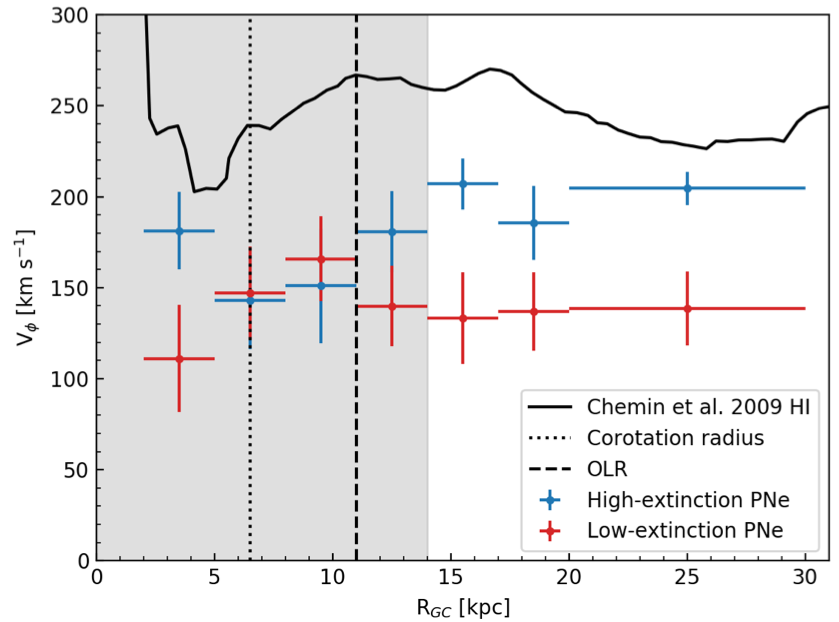}
	\caption{Rotational velocity for the high- and low-extinction PNe are shown in blue and red respectively. The black line shows the \ion{H}{i} rotation velocity from \citet{Chemin09}. The co-rotation radius (black dotted line) and outer Lindblad resonance (OLR; black dashed line) of the M31 bar are as found by the models of \citet{Blana18}. The grey shaded region is possibly influenced by different dynamical heating events and is not discussed here.}
	\label{fig:vrot}
\end{figure}

\section{Analysis}
\label{sect:ana}

\subsection{Classification of Planetary Nebulae based on extinction measurements}
\label{sect:extb}

The distribution of the M31 PNe extinction values (see Figure~\ref{fig:av_hist}) exhibits a sharp drop at A$\rm_V = 0.75$ mag, increases again at A$\rm_V =$1--1.25 mag, and drops off gradually at larger values of A$\rm_V$. Figure~\ref{fig:av_hist} also shows the distribution of the LMC PNe extinction values \citep{rp10}, shifted such that their peak (originally in the A$\rm_V =$ 0.75--1 mag bin) is coincident with the distribution of the M31 PNe extinction values (A$\rm_V =$ 0.25--0.5 mag bin). The shifted distribution of the LMC PNe extinction values also shows a sharp drop at A$\rm_V = 0.75$ mag and gradually falls off while that of the M31 disc PNe shows a secondary peak at A$\rm_V=$1--1.25 mag.
The distribution of M31 PNe extinction values around the first higher peak possibly results from an older parent stellar population (numerically more prevalent) spawning PNe having lower circumstellar extinction values (further discussions in Section~\ref{sect:age}), while the secondary peak at higher circumstellar extinction values would indicate the Presence of a younger parent stellar population.  

We thus classify M31 PNe with extinction values higher and lower than A$\rm_V = 0.75$ mag as high- and low-extinction PNe respectively. 
Our PNe sample is then divided into 145 high- and 268 low-extinction PNe, which are expected to be associated with younger and older parent stellar populations respectively. We note that using a different extinction value within A$\rm_V = 0.65-0.85$ mag range for the classification of the two PN populations has negligible effect on the rotation curves obtained in Section~\ref{sect:rot}. The high-extinction PNe classification is not biased by the LOS dust attenuation in M31 as per our investigation in Appendix~\ref{app:dust}. Figure~\ref{fig:spat_pos} shows the spatial distribution of the PNe in the M31 disc.

\subsection{Rotation curves}
\label{sect:rot}
For both \citetalias{san12} and Bh+19b PNe, the LOS velocities (LOSV) are obtained from full spectral fitting, resulting in an uncertainty of 3 km s$^{-1}$. The PNe are de-projected on to the galaxy plane based on the position angle (PA = 38$^\circ$) and inclination (i = $77^\circ$) of M31 in the planer disc approximation. They are then binned into seven elliptical bins (Figure~\ref{fig:spat_pos}) with the first six bins covering 3 kpc each starting at a deprojected major axis radius R$\rm_{GC}= 2$ kpc from the center of M31 and the final bin covering R$\rm_{GC}=$ 20--30 kpc. PNe observed outside R$\rm_{GC}=$ 30 kpc probably belong to the inner halo substructures, possibly the Northern Spur, and are hence not included in the analysis. The position of the PNe in each bin can be described using cylindrical coordinates, with the \textit{z} = 0 kpc plane as the local plane of the galaxy, \textit{r} = 0 kpc as the galactic center, and $\phi$ measured counterclockwise from the position angle of M31. The LOSV for the PNe, $V\rm_{LOS}$, in each bin is then fitted by the following equation:
\begin{equation}
\label{eq:vel}
    $$\rm V_{LOS} = V_{sys} + V_{\phi}~cos(\phi)~sin(i) + V_{R}~sin(\phi)~sin(i) + V_{err}$$
\end{equation}
where $V\rm_{sys}$ is the systemic velocity of M31, assumed to be $-309$ km s$^{-1}$ \citep{merrett06}; $V\rm_{\phi}$ is the rotational velocity in the plane of the galaxy; $V\rm_{R}$ is the radial streaming motion that can be inwards or outwards; $\rm i$ is the inclination of M31 mentioned previously and $V\rm_{err}= 3 $ km s$^{-1}$ is the uncertainty in measurement. LOSVs for the high- and low-extinction PNe are fitted separately in each elliptical bin using LMFIT \citep{lmfit} to obtain $V\rm_{\phi}$, $V\rm_{R}$ and $\rm\phi$ as the parameters describing the mean motion of the PNe populations in each bin. We note that $V\rm_{Z}$, the off-plane motion in the \textit{z} direction, is considered to be zero as no net off-plane motion is expected for PNe in the disc.

\begin{figure}[t]
	\centering
	\includegraphics[width=\columnwidth,angle=0]{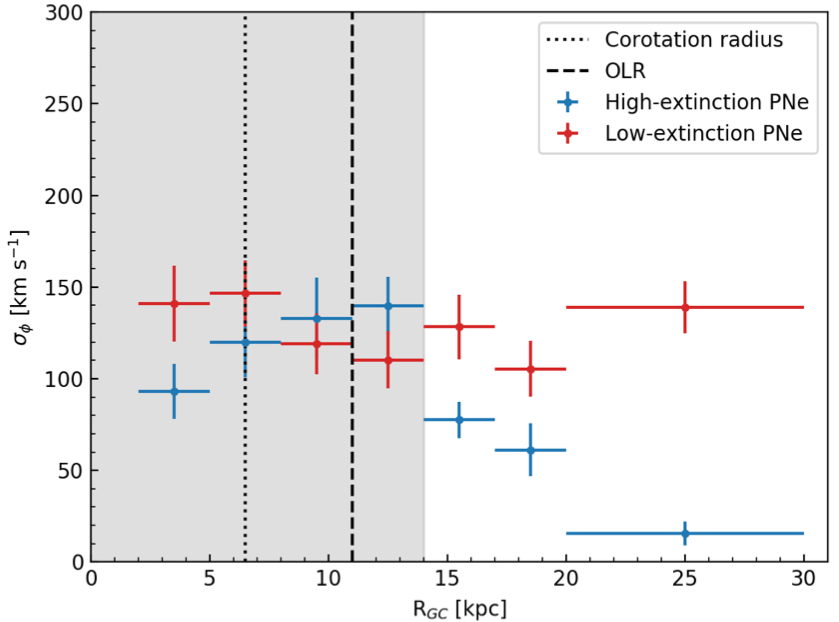}
	\caption{Rotational velocity dispersion for the high- and low-extinction PNe are known in blue and red respectively. The black lines and grey shaded region are same as in Figure~\ref{fig:vrot}. }
	\label{fig:sigrot}
\end{figure}

The obtained $V\rm_{\phi}$ rotation curves for the high- and low-extinction PNe are shown in blue and red respectively in Figure~\ref{fig:vrot}. The uncertainty in the fitted $V\rm_{R}$ is quite high and their values are close to zero in each bin. Thus, no clear evidence of radial streaming motion is found in either PNe population. Setting $V\rm_{R}=0$ km s$^{-1}$ also has negligible effect on the rotation curves. The difference in rotational velocities between the gas and the stellar population in a disc is a measure of the asymmetric drift and it is higher for older stellar populations which have more non-circular orbits as a result of dynamical heating \citep{Str46}. Outside R$\rm_{GC}= 14$ kpc, the high-extinction PNe have a rotational velocity closer to that of the \ion{H}{i} gas derived by \citet{Chemin09}, indicative of a dynamically young population, while that of the low-extinction PNe is much further away from that of the \ion{H}{i} gas, indicative of a dynamically older population \citep{Str46}. 

\begin{figure}[t]
	\centering
	\includegraphics[width=\columnwidth,angle=0]{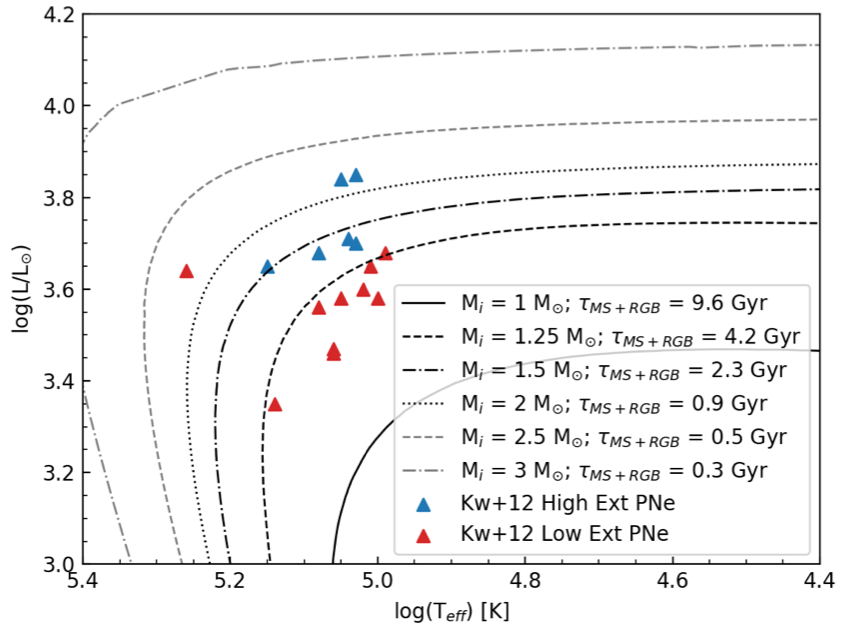}
	\caption{The high- and low-extinction PNe observed by \citetalias{Kw12} are shown in blue and red respectively in the $\rm log(L/L_{\odot})$ vs. $\rm log(T_{eff})$ plot. The stellar evolution tracks from \citet{MillerB16} corresponding to metallicity, $\rm Z_{0}=0.01$, are shown in black. The initial stellar mass and $\rm\tau_{MS+AGB}$ are also labeled.}
	\label{fig:kwit_ages}
\end{figure}

We estimate the rotational velocity dispersion, $\rm\sigma_{\phi}$, as the standard deviation with respect to the fitted $V\rm_{\phi}$ in each bin. The $\rm\sigma_{\phi}$ profiles for the high- and low-extinction PNe are shown in Figure~\ref{fig:sigrot}. Outside R$\rm_{GC}= 14$ kpc, $\rm\sigma_{\phi}$ is lower for the high-extinction PNe, as expected for a dynamically young population, than that measured for low-extinction PNe, a dynamically older population. $\rm\sigma_{\phi}$ for the low-extinction PNe population increases sharply in the outermost bin. This may be due to the presence of PNe associated with the M31 inner halo substructures like the Northern Spur or the NGC 205 loop at this distance from the M31 center. Within 14 kpc, both the high- and low-extinction PNe samples show an overall reversal in the $V\rm_{\phi}$ rotation curves and the $\rm\sigma_{\phi}$ but both populations are dynamically hot. While this maybe linked to the interaction of the disc with the bar in M31 as modelled by \citet{Blana18} for the inner two bins, other sources of dynamical heating may be at play for R$\rm_{GC}= 8-14$ kpc, either stemming just from the secular evolution of the disc and/or through a merger event. This will be investigated in a forthcoming paper (Bhattacharya et al. 2019b in preparation). Given the large values of $\rm\sigma_{\phi} \approx 130 $ km s$^{-1}$ for the low-extinction PNe, their parent stellar population may be distributed as a flattened spheroid, rather than a planer disc. Given the inclination of the M31 disc, deprojecting these PNe as a planer disc may result in an overestimate of their R$\rm_{GC}$ values, leading to a bias in the estimated $\rm\sigma_{\phi}$. We investigate the effect of disc thickness in Appendix~\ref{app:scale}. We find that the scale height of the low-extinction PNe is H$\rm_{Low~ext} \approx 0.86$ kpc. Within our 3 kpc bin sizes, only $\sim10\%$ of the low-extinction PNe may be included in a different bin. The effect on the estimated $\rm\sigma_{\phi}$ values of these $\sim10\%$ PNe in different bins is within the measurement uncertainties.


\subsection{Ages of the M31 disc Planetary Nebulae}
\label{sect:age}

\citetalias{Kw12} observed sixteen PNe in the outer disc of M31 to measure various emission lines and determine chemical abundances. They used the CLOUDY photoionization codes \citep{cloudy} to estimate the bolometric luminosity ($\rm L/L_{\odot}$) and effective temperature ($T\rm_{eff}$) of the central stars of these PNe. Figure~\ref{fig:kwit_ages} shows their estimated $\rm log(L/L_{\odot})$ vs. $\rm log(T_{eff})$, coloured by their extinction classification (high-extinction: blue; low-extinction: red). The post-AGB stellar evolution tracks from \citet{MillerB16} for a metallicity $Z\rm_{0}=0.01$ are also plotted in Figure~\ref{fig:kwit_ages}. It is clear that the high-extinction PNe in this sub-sample lie either around the tracks corresponding to an initial progenitor mass of 1.5 $M\rm_{\odot}$ and age ($\rm\tau_{MS+AGB}$; lifetime in main-sequence and AGB phases) of 2.3 Gyr or are even younger with higher initial progenitor masses. The low-extinction PNe in this sub-sample, barring one, are older than 4.2 Gyr with initial progenitor mass lower than 1.25 $M\rm_{\odot}$. We note that these ages maybe uncertain up to $\sim1$ Gyr based on the estimations by \citetalias{Kw12}. We may thus assign the mean ages corresponding to the \citetalias{Kw12} high- ($\sim2.5$ Gyr) and low- ($\sim4.5$ Gyr) extinction PNe to those with the corresponding extinction values in the \citetalias{san12} and Bh+19b PNe population.  

\begin{figure}[t]
	\centering
	\includegraphics[width=\columnwidth,angle=0]{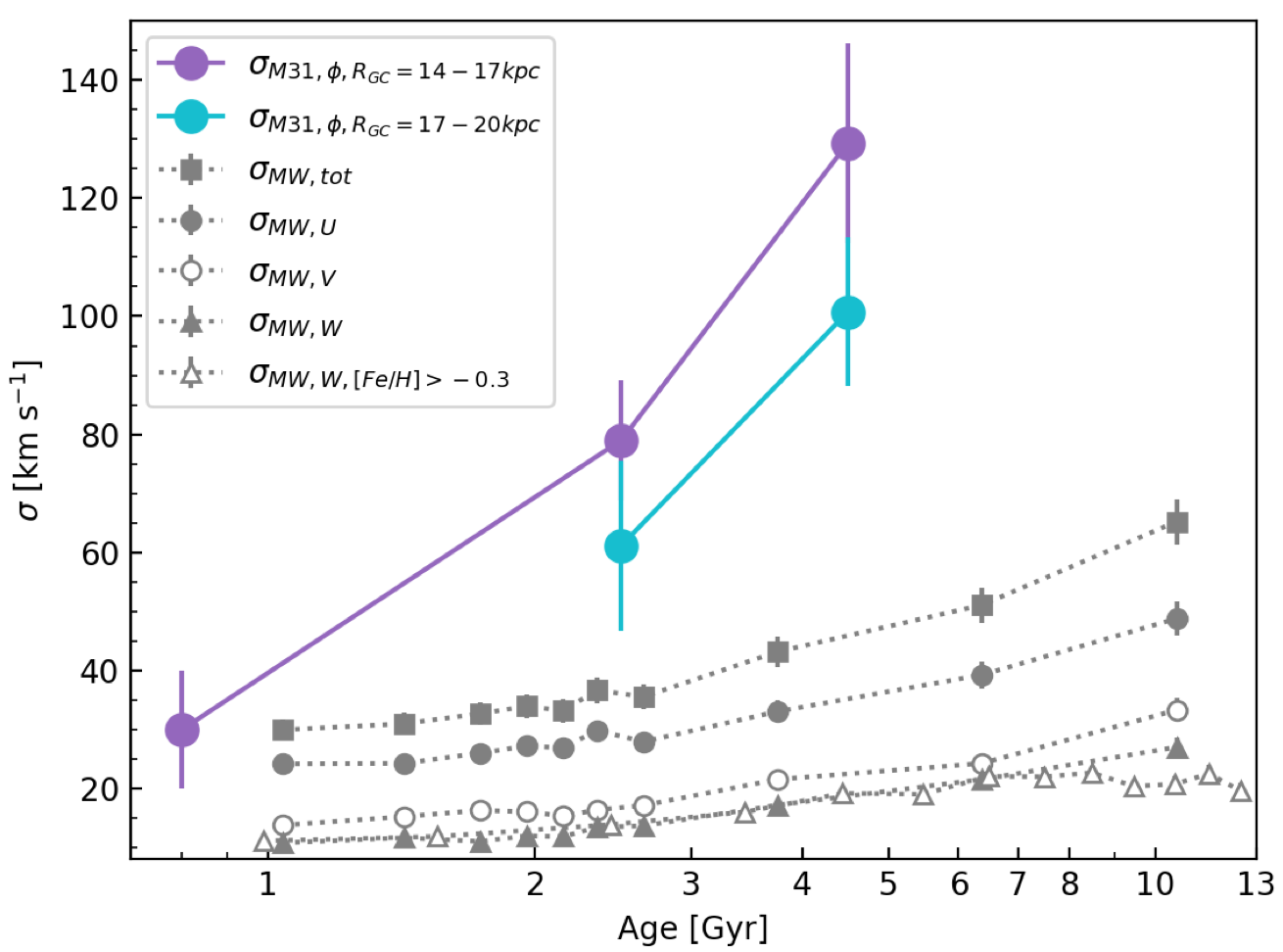}
	\caption{The AVR of PNe in the M31 disc at R$\rm_{GC}=$ 14--17 and 17--20 kpc is shown in magenta and cyan respectively. The assigned age is shown in log scale with the MS \citepalias[$\sim30$ Myr age;][]{dorman15} in the R$\rm_{GC}=$ 14--17 kpc bin shown at 0.8 Gyr for visual clarity. The AVR obtained in the solar neighbourhood in the MW \citep{Nord04} is shown in grey for comparison. Their total velocity dispersion is shown with squares while the velocity dispersion in the space velocity components (\textit{U, V, W}) are shown with filled circles, open circles and filled triangles respectively. We also present, with open triangles, the velocity dispersion in the W component from \citet{Aniyan18} for only those MW stars with [Fe/H]<-0.3, showing a flattening in the MW AVR at older ages. }
	\label{fig:age_vd_mw}
\end{figure}

\section{Age-velocity dispersion relation}
\label{sect:avr}

\subsection{The observed age-velocity dispersion relation in M31}
We obtain the AVR in M31 for two elliptical bins with R$\rm_{GC}=$ 14--17 and 17--20 kpc. They are presented in Figure~\ref{fig:age_vd_mw}, clearly showing the increase in the velocity dispersion with age. In Figure~\ref{fig:age_vd_mw}, we also present the age-velocity dispersion value for the MS ($\sim30$ Myr age), $\rm \sigma_{\phi,~MS}= 30\pm 10$ km s$^{-1}$, obtained by \citetalias{dorman15} in the R$\rm_{GC}=$ 14--17 kpc bin.

Based on models fitting the star formation rate, gas profiles and metallicity distributions of the M31 and MW discs, \citet{Yin09} find that R$\rm_{GC}$=19 kpc in the M31 disc is the equivalent distance, in disc scale lengths, of the Sun (R$\rm_{\odot}$=8 kpc) in the MW disc. We thus compare the velocity dispersion of the MW disc obtained in the solar neighbourhood by \citet{Nord04} to our $\rm\sigma_{\phi}$ in the R$\rm_{GC}$=17--20 kpc bin, where $\rm\sigma_{\phi,~2.5~Gyr}=\rm\sigma_{\phi,~High~ext}= 61\pm 14$ km s$^{-1}$ and $\rm\sigma_{\phi,~4.5~Gyr}=\rm\sigma_{\phi,~Low~ext}= 101\pm 13$ km s$^{-1}$. \citet{Nord04} describe the MW velocity dispersion in space-velocity components (\textit{U, V, W}), defined in a right-handed Galactic system with \textit{U} pointing towards the Galactic centre, \textit{V} in the direction of rotation, and \textit{W} towards the north Galactic pole (Figure~\ref{fig:age_vd_mw}). The equivalent in the MW disc for the $\rm\sigma_{\phi}$ in the M31 disc would be some combination of $\rm\sigma_{MW, U}$ and $\rm\sigma_{MW, V}$ with a value intermediate between the two \citepalias{dorman15}. We compare our obtained $\rm\sigma_{\phi}$ of the M31 disc with the $\rm\sigma_{MW, U}$, which is $\sim29$ km s$^{-1}$ and $\sim35$ km s$^{-1}$ for 2.5 Gyr and 4.5 Gyr old populations respectively. In the R$\rm_{GC}$=17--20 kpc bin, the $\rm\sigma_{\phi}$ of the 2.5 Gyr and 4.5 Gyr old population in M31 are about twice and thrice that of the $\rm\sigma_{MW, U}$ of the 2.5 Gyr and 4.5 Gyr old MW thin disc population respectively. 

\subsection{Comparison with previously measured and simulated age-velocity dispersion relations}
\label{sect:dis}

The AVR in M31 was previously estimated by \citetalias{dorman15} from the $\rm\sigma_{LOS}$ of stars whose classification in different age bins suffered from ambiguity. Their observations were also limited to the PHAT survey footprint, covering about a quarter of the M31 disc along its major axis out to R$\rm_{GC}\sim 18$ kpc. Our observed PNe sample covers the entire M31 disc out R$\rm_{GC}= 30$ kpc, and the high- and low-extinction PNe are well separated in age (Figure~\ref{fig:kwit_ages}). The $\rm\sigma_{\phi}$ values for the high- and low-extinction PNe agree within errors with that obtained by \citetalias{dorman15} for older AGB ($\sim 2$ Gyr old) and RGB ($\sim 4$ Gyr old) stars respectively. \citet{Quirk19} fitted the rotation curves for stellar populations identified by \citetalias{dorman15}. In the R$\rm_{GC}=14-17$ kpc bin, $V\rm_{\phi}$ for the high-extinction PNe is in good agreement with that obtained by \citet{Quirk19} for older AGB stars, but for the low-extinction PNe it is lower than that of RGB stars by $\sim 30$ km s$^{-1}$. This is possibly due to their RGB population being contaminated by younger AGB stars, resulting in a $V\rm_{\phi}$ value that is closer to that of the HI gas.
 
The AVR for the M31 disc shows a steep slope between 0--2.5 Gyr age range and an even steeper slope between 2.5--4.5 Gyr age range than those for the MW disc in similar age bins. The AVR of the MW disc is considered to be driven by secular evolution channels \citep[see review by][]{Sellwood14}. An AVR with velocity dispersion increasing gradually with age is also measured in simulated disc galaxies with similarly quiescent merger histories \citep[from zoom-in cosmological simulations by][]{House11, Martig14}. However, simulated disc galaxies undergoing a single merger show significant increase in the velocity dispersion for stellar populations older than the end of the merger \citep[][see their Figure 2]{Martig14}, with larger velocity dispersion for higher merger mass ratios. After the end of the merger, it takes $\sim2$ Gyr for stellar populations to form with velocity dispersion values similar to those for quiescent discs. The high $\rm\sigma_{\phi,~4.5~Gyr}$ values in the M31 disc is reminiscent of those seen in populations older than the merger event in simulated galaxies. The lower $\rm\sigma_{\phi,~2.5~Gyr}$ values in the M31 disc is reminiscent of the lower values predicted by simulations some time after the end of the merger. Finally, the velocity dispersion for the MS in M31 is then akin to that for quiescent discs, also observed at least $\sim2$ Gyr after the merger event in the simulated galaxies. Hence, we may deduce from the observed AVR in the M31 disc, that a single merger event took place 2.5 -- 4.5 Gyr ago.

\subsection{Estimation of the merger mass ratio}
\label{sect:ratio}

In the framework of a single merger in the M31 disc, we estimate the merger mass ratio and satellite mass required to produce the dynamically hot 4.5 Gyr old population with disc scale height  H$\rm_{4.5~Gyr} = H_{Low~ext} \approx 0.86$ kpc (Appendix~\ref{app:scale}). We utilize the relation between disc scale height (H) and satellite-to-disc-mass-ratio ($\rm M_{sat}/M_{disc}$) described by \citet{Hopkins08} for a satellite galaxy (assumed to be a rigid body) that merged with a disc galaxy (assumed to be a thin disc) on an in-plane prograde radial orbit. The relation in the case of a satellite merging with a \citet{Mestel63} disc galaxy, having constant circular velocity $V_{c,disc}$, is as follows: 
\begin{equation}
\label{eq:mass}
    $$\rm\frac{\Delta H}{R_{e,disc}} = \alpha_{H}~ (1-f_{gas}) ~\frac{M_{sat}}{M_{disc}}~\tilde{h}(R/R_{e,disc})$$
\end{equation}
where $\rm\Delta H$ gives the increase in scale height in the disc galaxy following the merger; $\rm\alpha_{H}= 1.6\tilde{v}$ is a derived constant with $\rm\tilde{v}=(V_{c,disc}/V_{h})^2$, $V_{h}$ being the halo circular velocity; $\rm f_{gas}$ is the gas fraction in both the disc galaxy and satellite (assumed to be equal) before the merger; $\rm R$ is the galactocentric radius of the population with scale height H ; and $\rm R_{e,disc}$ is the disc effective radius. 

We assume that the M31 disc evolved by secular evolution prior to the merger event. Hence we adopt the scale height of the old thin disc of the MW as measured in the solar neighbourhood, H$_{MW}\approx300$ pc \citep[see][and references therein]{bhg16}, as the pre-merger scale height $\rm H_{pre-merger}$ for the M31 disc. Thus,  $\rm H_{pre-merger} \approx H_{MW}\approx0.3$ kpc and $\rm\Delta H = H_{4.5~Gyr}- H_{pre-merger} \approx 0.56$ kpc. $\rm R=18.5$ is the median of the R$\rm_{GC}$=17--20 kpc bin, which is the equivalent disc scale lengths in M31 to the solar neighbourhood. From \citet{Blana18}, we adopt $V_{c,disc}=250$ km s$^{-1}$, $\rm R_{e,disc}=9.88$ and $V_{h}=182$ km s$^{-1}$. The present day gas fraction in M31 is $\sim9\%$ \citep{Yin09} but M31 is observed to have undergone a burst of star formation $\sim2$ Gyr ago which produced $\sim10\%$ of its mass \citep{wil17}. Assuming that the stellar mass formed in this burst was present as gas mass before the merger, we adopt $\rm f_{gas}=0.19$. Plugging these values into Equation~\ref{eq:mass}, we obtain $\rm M_{sat}/M_{disc} \approx 0.21$ or $\rm M_{sat}:M_{disc} \approx 1:5$. Given that the total mass of the M31 disc is $\rm 7 \times 10^{10} ~M_{\odot}$ \citep{Yin09}, a $\rm 1.4 \times 10^{10} ~M_{\odot}$ satellite is required to dynamically heat the M31 disc.

\section{Summary and conclusion}
\label{sect:sum}

We classify the observed sample of PNe based on their measured extinction values into high- and low-extinction PNe which are respectively associated with 2.5 Gyr and 4.5 Gyr parent populations. By fitting rotation curves to the two PNe populations in de-projected elliptical bins, we find that the high- and low-extinction PNe are dynamically colder and hotter respectively, especially  at R$\rm_{GC}=14-20$ kpc (Figures~\ref{fig:vrot},~\ref{fig:sigrot}). We thus obtain the AVR at these radii to find that $\rm\sigma_{\phi}$ increases with age in the M31 disc, which is dynamically much hotter than the stars in the MW disc of corresponding ages.  

There is an interesting timescale coincidence between the age of the high-extinction PNe and the $\sim2$ Gyr old burst of star formation observed both in the stellar disc and inner halo of M31 \citep{bernard15,wil17}. We speculate that most of the high-extinction PNe, causing the secondary peak in the extinction distribution (Figure~\ref{fig:av_hist}), are those whose progenitors formed during the $\sim2$ Gyr old star formation burst, while the low-extinction PNe were likely formed earlier. The high-extinction PNe are kinematically tracing the younger thin disc of M31 outside R$\rm_{GC}= 14$ kpc from the center. They are clearly separated, both in $\rm V_{\phi}$ and $\rm\sigma_{\phi}$, from the dynamically hotter low-extinction PNe which may be associated with the thicker disc. Some low-extinction PNe may also be associated with the old thin disc and inner halo of M31. 

 Using hydrodynamical simulations, \citet{ham18} argue that a single major merger 2 -- 3 Gyr ago, where the satellite eventually coalesced to build up the M31 bulge after multiple passages, can explain the dynamical heating of the M31 disc. They also predict a merger with mass ratio of at least 1:4.5 from their simulations, and also a decreasing trend in the velocity dispersion with radius, as observed, albeit within errors, in  Figure~\ref{fig:sigrot}. Such a merger could also explain the burst of star formation $\sim2$ Gyr ago and the presence of the M31 inner halo substructures. \citet{Fardal13} also use hydrodynamical simulations to predict the formation of the giant stream from a merger $\sim1$ Gyr ago with a $\rm \sim 3.2 \times 10^{9} M_{\odot}$ satellite. The AVR measured in the M31 disc using PNe is indicative of a single merger occurring 2.5 -- 4.5 Gyr ago with a merger mass ratio $\approx$ 1:5, with a $\rm 1.4 \times 10^{10} ~M_{\odot}$ satellite galaxy. Such a galaxy would have been the third largest member of the local group, more massive than M33 \citep{Kam17}. This is consistent with the prediction from \citet{ham18}.  In conclusion, the kinematics of the M31 disc PNe have been able to shed light on the recent dynamical evolution of M31. Our next step is to utilize PNe to further investigate the interface of the disc and inner halo of M31.
\begin{acknowledgements}
      SB acknowledges support from the IMPRS on Astrophysics at the LMU Munich. We are grateful to the anonymous referee for the constructive comments that improved the manuscript. Based on observations obtained at the MMT Observatory, a joint facility of the Smithsonian Institution and the University of Arizona. Based on observations obtained with MegaPrime/MegaCam, a joint project of CFHT and CEA/DAPNIA, at the Canada-France-Hawaii Telescope (CFHT). This research made use of Astropy-- a community-developed core Python package for Astronomy \citep{Rob13}, Numpy \citep{numpy} and Matplotlib \citep{matplotlib}. This research also made use of NASA’s Astrophysics Data System (ADS\footnote{\url{https://ui.adsabs.harvard.edu}}).
\end{acknowledgements}


\bibliographystyle{aa} 
\bibliography{ref_pne.bib}

\begin{thebibliography}{52}
\expandafter\ifx\csname natexlab\endcsname\relax\def\natexlab#1{#1}\fi

\bibitem[{{Aniyan} {et~al.}(2018){Aniyan}, {Freeman}, {Arnaboldi}, {Gerhard},
  {Coccato}, {Fabricius}, {Kuijken}, {Merrifield}, \& {Ponomareva}}]{Aniyan18}
{Aniyan}, S., {Freeman}, K.~C., {Arnaboldi}, M., {et~al.} 2018, \mnras, 476,
  1909

\bibitem[{{Aniyan} {et~al.}(2016){Aniyan}, {Freeman}, {Gerhard}, {Arnaboldi},
  \& {Flynn}}]{Aniyan16}
{Aniyan}, S., {Freeman}, K.~C., {Gerhard}, O.~E., {Arnaboldi}, M., \& {Flynn},
  C. 2016, \mnras, 456, 1484

\bibitem[{{Astropy Collaboration} {et~al.}(2013){Astropy Collaboration},
  {Robitaille}, {Tollerud}, {Greenfield}, {Droettboom}, {Bray}, {Aldcroft},
  {Davis}, {Ginsburg}, {Price-Whelan}, {Kerzendorf}, {Conley}, {Crighton},
  {Barbary}, {Muna}, {Ferguson}, {Grollier}, {Parikh}, {Nair}, {Unther},
  {Deil}, {Woillez}, {Conseil}, {Kramer}, {Turner}, {Singer}, {Fox}, {Weaver},
  {Zabalza}, {Edwards}, {Azalee Bostroem}, {Burke}, {Casey}, {Crawford},
  {Dencheva}, {Ely}, {Jenness}, {Labrie}, {Lim}, {Pierfederici}, {Pontzen},
  {Ptak}, {Refsdal}, {Servillat}, \& {Streicher}}]{Rob13}
{Astropy Collaboration}, {Robitaille}, T.~P., {Tollerud}, E.~J., {et~al.} 2013,
  \aap, 558, A33

\bibitem[{{Bernard} {et~al.}(2015){Bernard}, {Ferguson}, {Chapman}, {Ibata},
  {Irwin}, {Lewis}, \& {McConnachie}}]{bernard15}
{Bernard}, E.~J., {Ferguson}, A. M.~N., {Chapman}, S.~C., {et~al.} 2015,
  \mnras, 453, L113

\bibitem[{{Bhattacharya} {et~al.}(2019){Bhattacharya}, {Arnaboldi}, {Hartke},
  {Gerhard}, {Comte}, {McConnachie}, \& {Caldwell}}]{Bh+19}
{Bhattacharya}, S., {Arnaboldi}, M., {Hartke}, J., {et~al.} 2019, A\&A, 624,
  A132

\bibitem[{{Bla{\~n}a} {et~al.}(2018){Bla{\~n}a}, {Gerhard}, {Wegg}, {Portail},
  {Opitsch}, {Saglia}, {Fabricius}, {Erwin}, \& {Bender}}]{Blana18}
{Bla{\~n}a}, M., {Gerhard}, O., {Wegg}, C., {et~al.} 2018, \mnras, 481, 3210

\bibitem[{{Bland-Hawthorn} \& {Gerhard}(2016)}]{bhg16}
{Bland-Hawthorn}, J. \& {Gerhard}, O. 2016, Annual Review of Astronomy and
  Astrophysics, 54, 529

\bibitem[{{Casagrande} {et~al.}(2011){Casagrande}, {Sch{\"o}nrich}, {Asplund},
  {Cassisi}, {Ram{\'\i}rez}, {Mel{\'e}ndez}, {Bensby}, \&
  {Feltzing}}]{Casagrande11}
{Casagrande}, L., {Sch{\"o}nrich}, R., {Asplund}, M., {et~al.} 2011, \aap, 530,
  A138

\bibitem[{{Chemin} {et~al.}(2009){Chemin}, {Carignan}, \& {Foster}}]{Chemin09}
{Chemin}, L., {Carignan}, C., \& {Foster}, T. 2009, \apj, 705, 1395

\bibitem[{{Ciardullo} \& {Jacoby}(1999)}]{Ciardullo99}
{Ciardullo}, R. \& {Jacoby}, G.~H. 1999, \apj, 515, 191

\bibitem[{{Coccato} {et~al.}(2009){Coccato}, {Gerhard}, {Arnaboldi}, {Das},
  {Douglas}, {Kuijken}, {Merrifield}, {Napolitano}, {Noordermeer},
  {Romanowsky}, {Capaccioli}, {Cortesi}, {De Lorenzi}, \& {Freeman}}]{coc09}
{Coccato}, L., {Gerhard}, O., {Arnaboldi}, M., {et~al.} 2009, \mnras, 394, 1249

\bibitem[{{Comer{\'o}n} {et~al.}(2019){Comer{\'o}n}, {Salo}, {Knapen}, \&
  {Peletier}}]{Comeron19}
{Comer{\'o}n}, S., {Salo}, H., {Knapen}, J.~H., \& {Peletier}, R.~F. 2019,
  \aap, 623, A89

\bibitem[{{Cortesi} {et~al.}(2013){Cortesi}, {Arnaboldi}, {Coccato},
  {Merrifield}, {Gerhard}, {Bamford}, {Romanowsky}, {Napolitano}, {Douglas},
  {Kuijken}, {Capaccioli}, {Freeman}, {Chies-Santos}, \& {Pota}}]{cor13}
{Cortesi}, A., {Arnaboldi}, M., {Coccato}, L., {et~al.} 2013, \aap, 549, A115

\bibitem[{{Dalcanton} {et~al.}(2015){Dalcanton}, {Fouesneau}, {Hogg}, {Lang},
  {Leroy}, {Gordon}, {Sand strom}, {Weisz}, {Williams}, \& {Bell}}]{Dal15}
{Dalcanton}, J.~J., {Fouesneau}, M., {Hogg}, D.~W., {et~al.} 2015, \apj, 814, 3

\bibitem[{{Dalcanton} {et~al.}(2012){Dalcanton}, {Williams}, {Lang}, {Lauer},
  {Kalirai}, {Seth}, {Dolphin}, {Rosenfield}, {Weisz}, {Bell}, {Bianchi},
  {Boyer}, {Caldwell}, {Dong}, {Dorman}, {Gilbert}, {Girardi}, {Gogarten},
  {Gordon}, {Guhathakurta}, {Hodge}, {Holtzman}, {Johnson}, {Larsen}, {Lewis},
  {Melbourne}, {Olsen}, {Rix}, {Rosema}, {Saha}, {Sarajedini}, {Skillman}, \&
  {Stanek}}]{dal12}
{Dalcanton}, J.~J., {Williams}, B.~F., {Lang}, D., {et~al.} 2012, The
  Astrophysical Journal Supplement Series, 200, 18

\bibitem[{{Delhaye}(1965)}]{Delhaye65}
{Delhaye}, J. 1965, {Solar Motion and Velocity Distribution of Common Stars}
  (University of Chicago Press, Chicago, ILL USA), 61

\bibitem[{{Dorman} {et~al.}(2015){Dorman}, {Guhathakurta}, {Seth}, {Weisz},
  {Bell}, {Dalcanton}, {Gilbert}, {Hamren}, {Lewis}, {Skillman}, {Toloba}, \&
  {Williams}}]{dorman15}
{Dorman}, C.~E., {Guhathakurta}, P., {Seth}, A.~C., {et~al.} 2015, \apj, 803,
  24

\bibitem[{{Draine} {et~al.}(2014){Draine}, {Aniano}, {Krause}, {Groves},
  {Sandstrom}, {Braun}, {Leroy}, {Klaas}, {Linz}, {Rix}, {Schinnerer},
  {Schmiedeke}, \& {Walter}}]{Draine14}
{Draine}, B.~T., {Aniano}, G., {Krause}, O., {et~al.} 2014, \apj, 780, 172

\bibitem[{{Fabricant} {et~al.}(2005){Fabricant}, {Fata}, {Roll}, {Hertz},
  {Caldwell}, {Gauron}, {Geary}, {McLeod}, {Szentgyorgyi}, {Zajac}, {Kurtz},
  {Barberis}, {Bergner}, {Brown}, {Conroy}, {Eng}, {Geller}, {Goddard},
  {Honsa}, {Mueller}, {Mink}, {Ordway}, {Tokarz}, {Woods}, {Wyatt}, {Epps}, \&
  {Dell'Antonio}}]{fab05}
{Fabricant}, D., {Fata}, R., {Roll}, J., {et~al.} 2005, Publications of the
  Astronomical Society of the Pacific, 117, 1411

\bibitem[{{Fardal} {et~al.}(2013){Fardal}, {Weinberg}, {Babul}, {Irwin},
  {Guhathakurta}, {Gilbert}, {Ferguson}, {Ibata}, {Lewis}, {Tanvir}, \&
  {Huxor}}]{Fardal13}
{Fardal}, M.~A., {Weinberg}, M.~D., {Babul}, A., {et~al.} 2013, \mnras, 434,
  2779

\bibitem[{{Ferland} {et~al.}(1998){Ferland}, {Korista}, {Verner}, {Ferguson},
  {Kingdon}, \& {Verner}}]{cloudy}
{Ferland}, G.~J., {Korista}, K.~T., {Verner}, D.~A., {et~al.} 1998,
  Publications of the Astronomical Society of the Pacific, 110, 761

\bibitem[{{Gilmore} \& {Reid}(1983)}]{Gilmore83}
{Gilmore}, G. \& {Reid}, N. 1983, \mnras, 202, 1025

\bibitem[{{Guhathakurta} {et~al.}(2005){Guhathakurta}, {Ostheimer}, {Gilbert},
  {Rich}, {Majewski}, {Kalirai}, {Reitzel}, \& {Patterson}}]{Guha05}
{Guhathakurta}, P., {Ostheimer}, J.~C., {Gilbert}, K.~M., {et~al.} 2005, arXiv
  e-prints [astro-ph/0502366]

\bibitem[{{Guhathakurta} {et~al.}(2006){Guhathakurta}, {Rich}, {Reitzel},
  {Cooper}, {Gilbert}, {Majewski}, {Ostheimer}, {Geha}, {Johnston}, \&
  {Patterson}}]{Guha06}
{Guhathakurta}, P., {Rich}, R.~M., {Reitzel}, D.~B., {et~al.} 2006, \aj, 131,
  2497

\bibitem[{{Hammer} {et~al.}(2018){Hammer}, {Yang}, {Wang}, {Ibata}, {Flores},
  \& {Puech}}]{ham18}
{Hammer}, F., {Yang}, Y.~B., {Wang}, J.~L., {et~al.} 2018, \mnras, 475, 2754

\bibitem[{{Hopkins} {et~al.}(2008){Hopkins}, {Hernquist}, {Cox}, {Younger}, \&
  {Besla}}]{Hopkins08}
{Hopkins}, P.~F., {Hernquist}, L., {Cox}, T.~J., {Younger}, J.~D., \& {Besla},
  G. 2008, \apj, 688, 757

\bibitem[{{House} {et~al.}(2011){House}, {Brook}, {Gibson},
  {S{\'a}nchez-Bl{\'a}zquez}, {Courty}, {Few}, {Governato}, {Kawata},
  {Ro{\v{s}}kar}, {Steinmetz}, {Stinson}, \& {Teyssier}}]{House11}
{House}, E.~L., {Brook}, C.~B., {Gibson}, B.~K., {et~al.} 2011, \mnras, 415,
  2652

\bibitem[{Hunter(2007)}]{matplotlib}
Hunter, J.~D. 2007, Computing In Science \& Engineering, 9, 90

\bibitem[{{Kam} {et~al.}(2017){Kam}, {Carignan}, {Chemin}, {Foster}, {Elson},
  \& {Jarrett}}]{Kam17}
{Kam}, S.~Z., {Carignan}, C., {Chemin}, L., {et~al.} 2017, \aj, 154, 41

\bibitem[{{Kwitter} {et~al.}(2012){Kwitter}, {Lehman}, {Balick}, \&
  {Henry}}]{Kw12}
{Kwitter}, K.~B., {Lehman}, E. M.~M., {Balick}, B., \& {Henry}, R.~B.~C. 2012,
  \apj, 753, 12

\bibitem[{{Martig} {et~al.}(2014){Martig}, {Minchev}, \& {Flynn}}]{Martig14}
{Martig}, M., {Minchev}, I., \& {Flynn}, C. 2014, \mnras, 443, 2452

\bibitem[{{McConnachie} {et~al.}(2018){McConnachie}, {Ibata}, {Martin},
  {Ferguson}, {Collins}, {Gwyn}, {Irwin}, {Lewis}, {Mackey}, {Davidge},
  {Arias}, {Conn}, {C{\^o}t{\'e}}, {Crnojevic}, {Huxor}, {Penarrubia},
  {Spengler}, {Tanvir}, {Valls-Gabaud}, {Babul}, {Barmby}, {Bate}, {Bernard},
  {Chapman}, {Dotter}, {Harris}, {McMonigal}, {Navarro}, {Puzia}, {Rich},
  {Thomas}, \& {Widrow}}]{mcc18}
{McConnachie}, A.~W., {Ibata}, R., {Martin}, N., {et~al.} 2018, \apj, 868, 55

\bibitem[{{Merrett} {et~al.}(2006){Merrett}, {Merrifield}, {Douglas},
  {Kuijken}, {Romanowsky}, {Napolitano}, {Arnaboldi}, {Capaccioli}, {Freeman},
  {Gerhard}, {Coccato}, {Carter}, {Evans}, {Wilkinson}, {Halliday}, \&
  {Bridges}}]{merrett06}
{Merrett}, H.~R., {Merrifield}, M.~R., {Douglas}, N.~G., {et~al.} 2006, \mnras,
  369, 120

\bibitem[{{Mestel}(1963)}]{Mestel63}
{Mestel}, L. 1963, \mnras, 126, 553

\bibitem[{{Miller Bertolami}(2016)}]{MillerB16}
{Miller Bertolami}, M.~M. 2016, \aap, 588, A25

\bibitem[{{Newville} {et~al.}(2014){Newville}, {Stensitzki}, {Allen}, \&
  {Ingargiola}}]{lmfit}
{Newville}, M., {Stensitzki}, T., {Allen}, D.~B., \& {Ingargiola}, A. 2014,
  {LMFIT: Non-Linear Least-Square Minimization and Curve-Fitting for Python}

\bibitem[{{Nordstr{\"o}m} {et~al.}(2004){Nordstr{\"o}m}, {Mayor}, {Andersen},
  {Holmberg}, {Pont}, {J{\o}rgensen}, {Olsen}, {Udry}, \& {Mowlavi}}]{Nord04}
{Nordstr{\"o}m}, B., {Mayor}, M., {Andersen}, J., {et~al.} 2004, \aap, 418, 989

\bibitem[{Oliphant(2015)}]{numpy}
Oliphant, T.~E. 2015, Guide to NumPy, 2nd edn. (USA: CreateSpace Independent
  Publishing Platform)

\bibitem[{{Pulsoni} {et~al.}(2018){Pulsoni}, {Gerhard}, {Arnaboldi}, {Coccato},
  {Longobardi}, {Napolitano}, {Moylan}, {Narayan}, {Gupta}, {Burkert},
  {Capaccioli}, {Chies-Santos}, {Cortesi}, {Freeman}, {Kuijken}, {Merrifield},
  {Romanowsky}, \& {Tortora}}]{pul18}
{Pulsoni}, C., {Gerhard}, O., {Arnaboldi}, M., {et~al.} 2018, \aap, 618, A94

\bibitem[{{Quinn} \& {Goodman}(1986)}]{Quinn86}
{Quinn}, P.~J. \& {Goodman}, J. 1986, \apj, 309, 472

\bibitem[{{Quirk} {et~al.}(2019){Quirk}, {Guhathakurta}, {Chemin}, {Dorman},
  {Gilbert}, {Seth}, {Williams}, \& {Dalcanton}}]{Quirk19}
{Quirk}, A., {Guhathakurta}, P., {Chemin}, L., {et~al.} 2019, \apj, 871, 11

\bibitem[{{Reid} \& {Parker}(2010)}]{rp10}
{Reid}, W.~A. \& {Parker}, Q.~A. 2010, \mnras, 405, 1349

\bibitem[{{Roberts}(1966)}]{Roberts66}
{Roberts}, M.~S. 1966, \apj, 144, 639

\bibitem[{{Sanders} {et~al.}(2012){Sanders}, {Caldwell}, {McDowell}, \&
  {Harding}}]{san12}
{Sanders}, N.~E., {Caldwell}, N., {McDowell}, J., \& {Harding}, P. 2012, \apj,
  758, 133

\bibitem[{{Sellwood}(2014)}]{Sellwood14}
{Sellwood}, J.~A. 2014, Reviews of Modern Physics, 86, 1

\bibitem[{{Str{\"o}mberg}(1925)}]{Stromberg25}
{Str{\"o}mberg}, G. 1925, \apj, 61, 363

\bibitem[{{Str{\"o}mberg}(1946)}]{Str46}
{Str{\"o}mberg}, G. 1946, \apj, 104, 12

\bibitem[{{Ventura} {et~al.}(2014){Ventura}, {Dell'Agli}, {Schneider}, {Di
  Criscienzo}, {Rossi}, {La Franca}, {Gallerani}, \& {Valiante}}]{Ventura14}
{Ventura}, P., {Dell'Agli}, F., {Schneider}, R., {et~al.} 2014, \mnras, 439,
  977

\bibitem[{{Wielen}(1977)}]{Wielen77}
{Wielen}, R. 1977, \aap, 60, 263

\bibitem[{{Williams} {et~al.}(2017){Williams}, {Dolphin}, {Dalcanton}, {Weisz},
  {Bell}, {Lewis}, {Rosenfield}, {Choi}, {Skillman}, \& {Monachesi}}]{wil17}
{Williams}, B.~F., {Dolphin}, A.~E., {Dalcanton}, J.~J., {et~al.} 2017, \apj,
  846, 145

\bibitem[{{Yin} {et~al.}(2009){Yin}, {Hou}, {Prantzos}, {Boissier}, {Chang},
  {Shen}, \& {Zhang}}]{Yin09}
{Yin}, J., {Hou}, J.~L., {Prantzos}, N., {et~al.} 2009, \aap, 505, 497

\bibitem[{{Yoachim} \& {Dalcanton}(2006)}]{Yoachim06}
{Yoachim}, P. \& {Dalcanton}, J.~J. 2006, \aj, 131, 226

\end{thebibliography}

\begin{appendix} 

\section{Effect of line-of-sight dust attenuation in extinction-based selection of Planetary Nebulae}
\label{app:dust}

\begin{figure*}[t]
	\centering
	\includegraphics[width=\linewidth,angle=0]{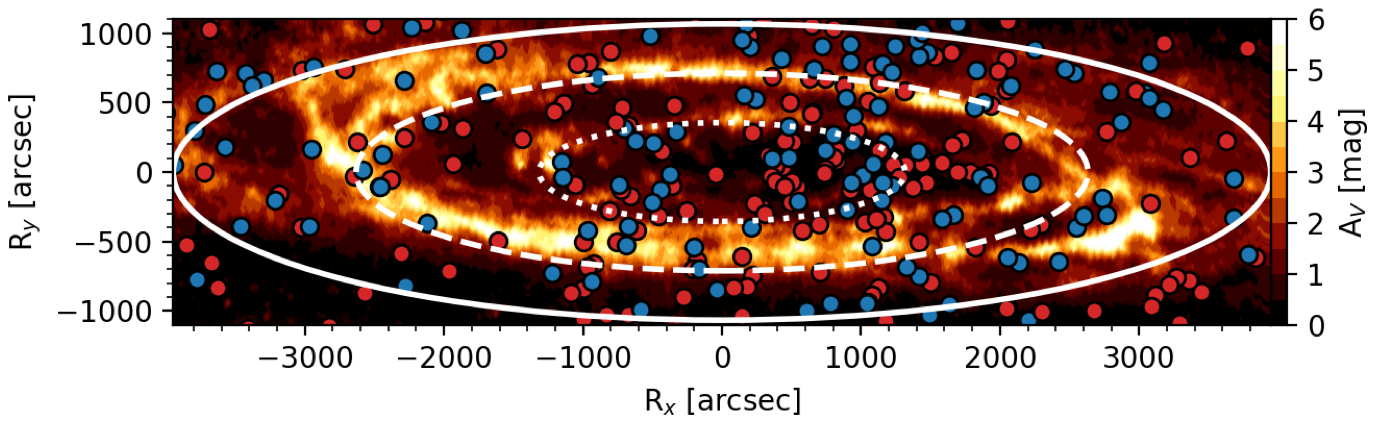}
	\caption{The high- (blue) and low- (red) extinction PNe are shown overlaid on the de-projected LOS extinction (A$\rm_V$) map of the inner region of M31 from \citet{Blana18}. The elliptical rings show 5 (dotted), 10 (dashed) and 15 (solid) kpc de-projected distances from the centre of M31. Both the high- and low-extinction PNe are not spatially correlated with the distribution of LOS extinction in the M31 disc.}
	\label{fig:dust}
\end{figure*}

\begin{figure}[t]
	\centering
	\includegraphics[width=\columnwidth,angle=0]{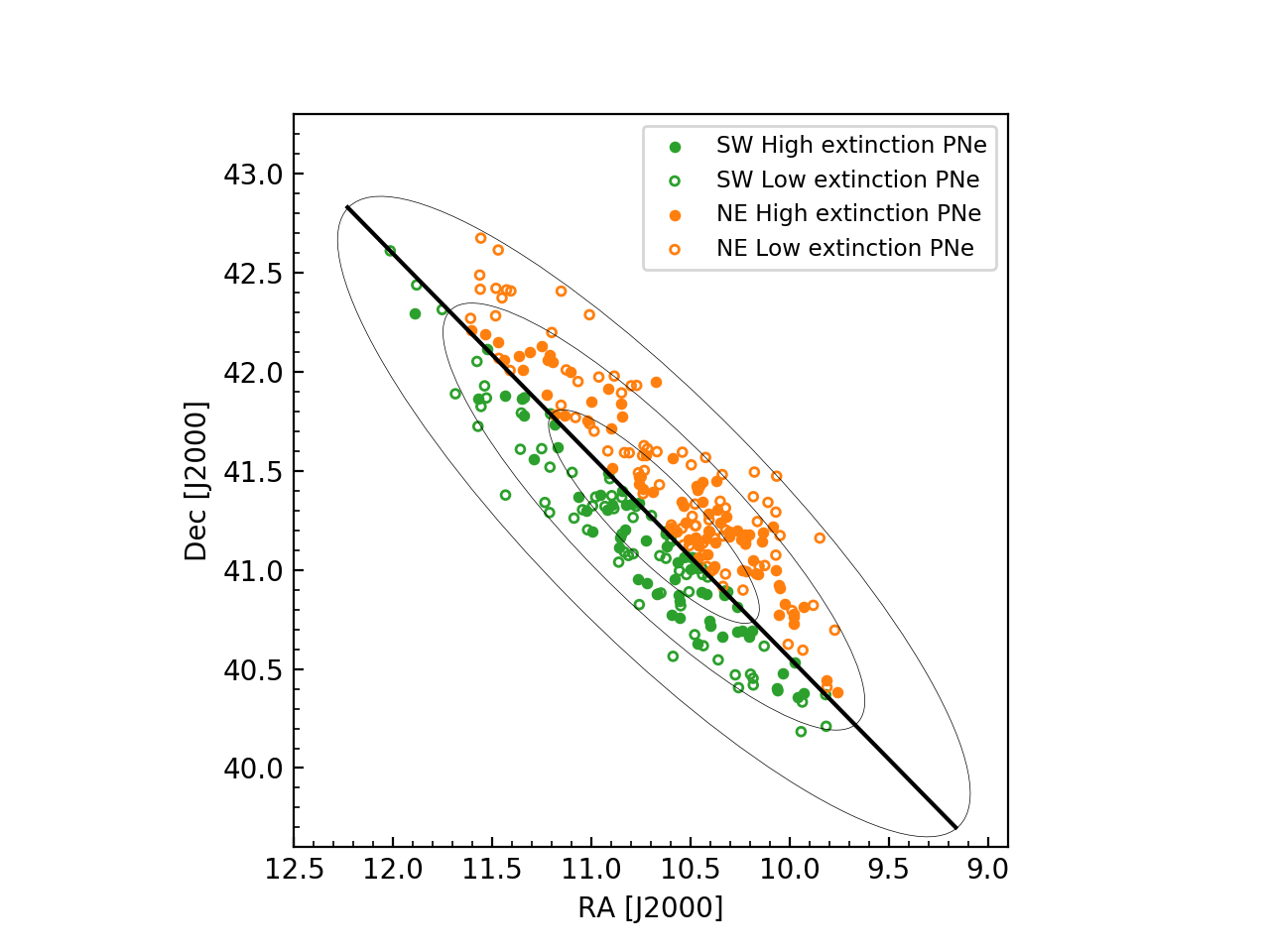}
	\caption{Position in space of high- and low-extinction PNe marked with filled and open circles respectively. The black line indicates the major axis of M31, dividing the PNe into the North-Eastern (NE) and South-Western (SW) halves of the M31 disc shown in orange and green respectively. The black ellipses show R$\rm_{GC}=$ 10 kpc (inner), 20 kpc (middle) and 30 kpc (outer). North is up, east is right.}
	\label{fig:spat_ext}
\end{figure}

\begin{figure}[t]
	\centering
	\includegraphics[width=\columnwidth,angle=0]{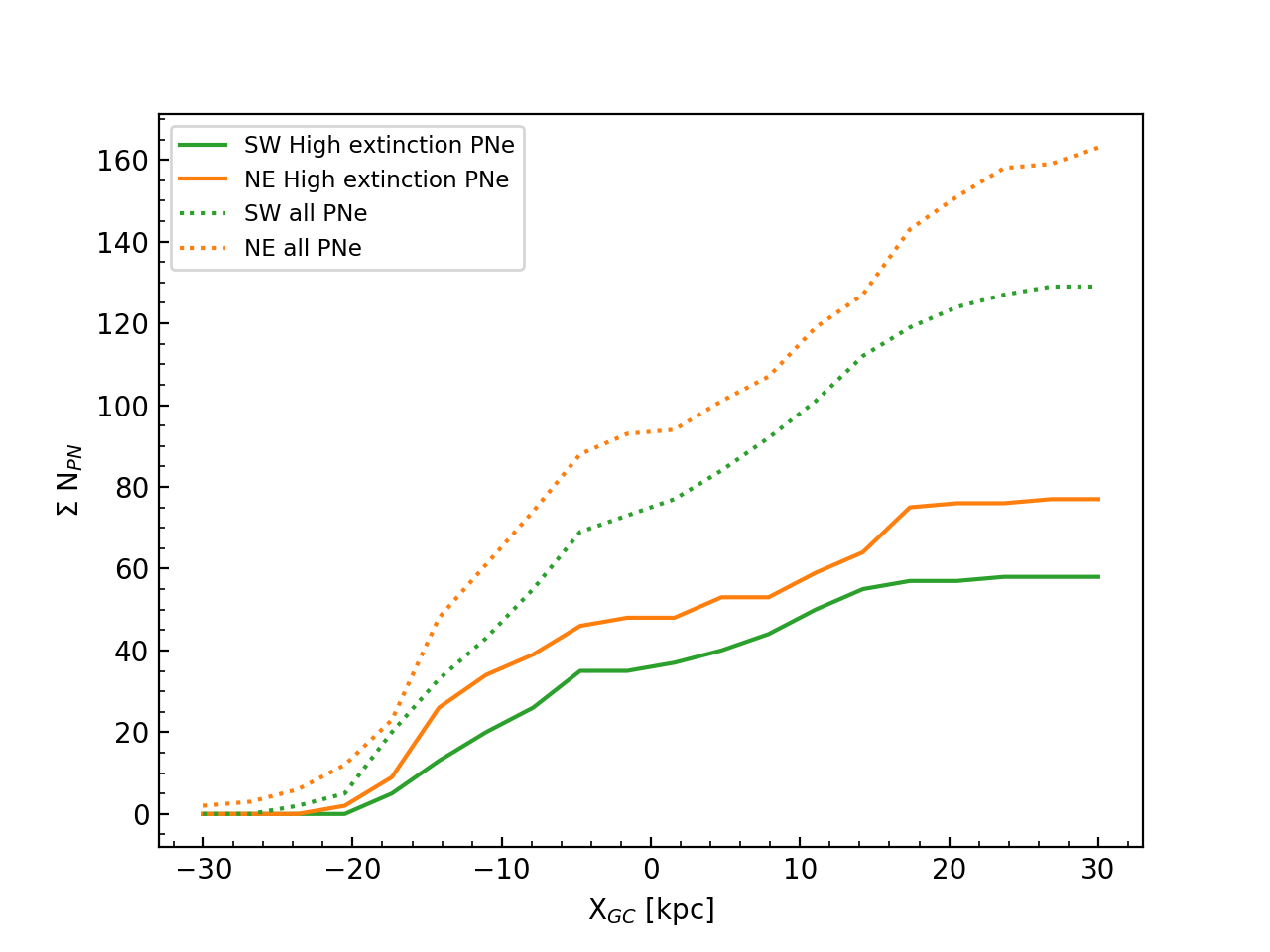}
	\caption{The cumulative sum of high extinction PNe (solid lines) and all PNe (dotted lines) for the NE (orange) and SW (green) along the deprojected major-axis distances, X$_{GC}$. Southern side of M31 is assigned negative values.}
	\label{fig:cum_ext}
\end{figure}

The PNe are separated into high and low extinction samples considering that the measured extinction values for high extinction PNe are dominated by their circumstellar extinction, with LOS dust attenuation equally affecting both samples. Any spatial correlation between higher LOS dust attenuation in the galaxy and the position of high-extinction PNe would result in PNe with low circumstellar extinction being misclassified. The dust in the M31 disc and IGM was mapped by \citet{Draine14} using near-infrared data from the Spitzer Space Telescope. This was utilized by \citet{Blana18} to obtain the LOS extinction map of the inner disc of M31 shown in Figure~\ref{fig:dust}, also showing the de-projected positions of the high- (blue) and low- (red) extinction PNe. It is clear that there is no preferential selection of the high-extinction PNe at the regions of high LOS extinction in the disc. Both the high- and low-extinction PNe are not spatially correlated with the distribution of LOS extinction. 

To further investigate this issue, we check whether the high-extinction PNe are preferentially found in the dustier regions of the M31 disc by taking in to account the geometry of its dust attenuation. The North East (NE) half of the M31 disc has higher LOS dust attenuation than the South West (SW) half \citep{Draine14}. We thus divide the PNe sample into two halves corresponding to these two regions of the M31 disc, as shown in Figure~\ref{fig:spat_ext}. The cumulative number of PNe is obtained along their deprojected major-axis distances (X$\rm_{PC}$), with the negative values assigned to the southern side of M31, for the high-extinction PNe (solid) and all PNe (dashed), for both the NE (orange) and SW (green) halves for the M31 disc (Figure~\ref{fig:cum_ext}). The cumulative distribution of both high-extinction PNe and the entire PNe sample clearly follow a similar distribution in both halves of the M31 disc with a slightly higher number of PNe observed in the NE half. A KS test shows that the high-extinction PNe in both halves follow the same distribution with a pvalue=0.99. 
We classify 2\% more PNe as high extinction in the SW half of M31 which is anti-correlated with the expected effects from the LOS dust attenuation. We thus conclude that the LOS dust attenuation effects do not drive the extinction-based classification of our PN populations.

\section{Validity of the planar disc assumption for the low-extinction Planetary Nebulae}
\label{app:scale}
At the inclination of M31, deprojecting the low-extinction PNe at high \textit{z} on a planer disc may result in assigned R$\rm_{GC}$ values which are different from the true ones. We first estimate the scale height of the low-extinction PNe (H$\rm_{Low~ext}$). In order to establish an extinction map of M31 RGB stars found by the PHAT survey, \citet{Dal15} modelled the geometry of the M31 thicker disc to describe the distribution of RGB \citepalias[$\sim 4$ Gyr old;][]{dorman15} and red clump stars in low-extinction regions. They found that this thicker disc has a ratio of vertical to horizontal exponential scale heights of $\rm h_z/h_r = 0.15$. From IRAC 3.6 $\mu$m band images of M31, \citet{Blana18} find that the M31 disc scale length $\rm h_r=5.71\pm0.08$ kpc. Thus, $\rm h_z=0.86 \pm 0.01$ for the M31 thicker disc. Since the RGB stars have mean age and kinematics close to that of the low-extinction PNe (see Sections~\ref{sect:age} \& \ref{sect:dis} for details), H$\rm_{Low~ext}=h_z\approx 0.86$ kpc.

Given the inclination of M31, PNe at \textit{z}$\sim 0.86$ kpc may be deprojected with an estimated R$_{GC}$ that is $\sim$0.2 kpc different from its true R$_{GC}$ value. For an exponentially decreasing stellar density ($\rho$) profile of the low-extinction PNe with \textit{z}, $\rho \propto e^{-\frac{z}{H\rm_{Low~ext}}}$, 1/4 of the low-extinction PNe in any bin may lie at \textit{z}$\sim 0.86$ kpc. Given the 3 kpc bin sizes used to determine $\rm\sigma_{\phi, Low~ext}$, only $\sim10\%$ of the low-extinction PNe may be included in a different bin. The measured $\rm\sigma_{\phi}$ values have $\sim10\%$ error, as in Figure~\ref{fig:sigrot}. Thus the planar disc assumption for PNe does not bias the $V\rm_{\phi}$ and $\rm\sigma_{\phi}$ profiles within the estimated errors.

\end{appendix}
\end{document}